\newif\ifeprint
\eprinttrue
\def\figuredir{.}

\newlength\xx
\ifeprint

\documentclass
[rmp,twocolumn,amsfonts,amssymb,amsmath,showkeys,letterpaper,
raggedbottom,floatfix]{revtex4}
\makeatletter\def\raggedcolumn@skip{\vskip\z@\@plus.0001fil\relax}\makeatother
\def\eprint#1{E-print \urlalt{http://arxiv.org/abs/#1}{arXiv:#1}}
\def\;{\!}

\else

\documentclass[final,twocolumn,natbib]{svjour3}
\journalname{J. Geod.}
\usepackage{amsmath}
\usepackage{amssymb}
\usepackage{mathptmx}
\def\eprint#1{preprint \url{http://arxiv.org/abs/#1}}
\let\;\empty

\fi

\usepackage{array}
\date{September 21, 2011; revised \today}

\def\TITLE{Algorithms for geodesics}
\usepackage{ifpdf}
\usepackage{times}
\usepackage[bookmarks=false]{hyperref}
\usepackage{breakurl}

\hypersetup{pdftitle={\TITLE},pdfauthor={C. F. F. Karney}}
\usepackage{graphicx}

\def\d{\mathrm d}
\newcount\minutes \newcount\hours
\hours=\time \divide\hours by 60 \multiply\hours by 60
\minutes=\time \advance\minutes by -\hours \divide\hours by 60

\def\twodigit#1{\ifnum#1<10 0\fi\the#1}

\def\signsp{\setbox0\hbox{$-$}\setbox1\hbox{$0$}\dimen0\wd0%
\advance\dimen0-\wd1\setbox0\null\wd0=\dimen0\box0\relax}

\ifpdf\def\urlalt#1#2{\burlalt{#2}{#1}}\else\let\urlalt=\burlalt\fi
\def\doi#1{\urlalt{http://dx.doi.org/#1}{doi:#1}}
\def\tfrac#1/#2 {{\textstyle\frac{#1}{#2}}}
\def\v#1{\mathbf{#1}}
\newcommand{\ph}{\mathop{\mathrm{ph}}\nolimits}
\def\abs#1{\left|#1\right|}
\def\sqrta#1{\sqrt{\vphantom{\sin^2k^2}\smash{#1}}}
\def\dlmf#1#2{\urlalt{http://dlmf.nist.gov/#2}{#1}}
\newcommand{\sign}{\mathop{\mathrm{sign}}\nolimits}
\newcommand{\atantwo}{\mathop{\mathrm{atan2}}\nolimits}

\begin{document}

\ifeprint
\noindent\mbox{\begin{minipage}[b]{\textwidth}
\begin{flushright}
Link \url{http://geographiclib.sourceforge.net/geod.html}\par
\vspace{0.5ex}\eprint{1109.4448}\par
\vspace{2ex}
\end{flushright}
\end{minipage}
\hspace{-\textwidth}}
\fi

\title{\TITLE}
\ifeprint
\author{\href{http://charles.karney.info}{Charles F. F. Karney}}
\email{charles.karney@sri.com}
\affiliation{\href{http://www.sri.com}{SRI International},
201 Washington Rd, Princeton, NJ 08543-5300, USA}
\else
\author{Charles F. F. Karney}
\institute{C. F. F. Karney \at
\href{http://www.sri.com}{SRI International},
201 Washington Rd\\Princeton, NJ 08543-5300, USA\\\email{charles.karney@sri.com}
}
\fi

\ifeprint\else\maketitle\fi

\begin{abstract}

Algorithms for the computation of geodesics on an ellipsoid of
revolution are given.  These provide accurate, robust, and fast
solutions to the direct and inverse geodesic problems and they allow
differential and integral properties of geodesics to be computed.

\ifeprint
\else
\fi
\end{abstract}

\ifeprint\maketitle\fi

\section{Introduction}\label{intro}

\begin{figure}[tb]
\begin{center}
\includegraphics[scale=0.75,angle=0]{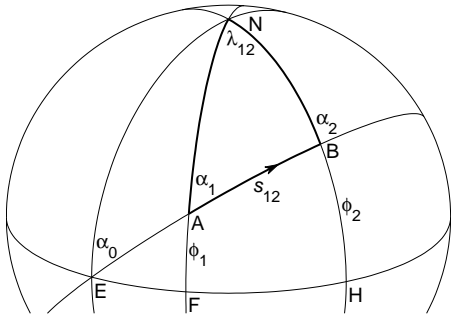}
\end{center}
\caption{\label{figtrig}
The ellipsoidal triangle $N\;AB$.  $N$ is the north pole, $N\;AF$ and
$N\;BH$ are meridians, and $AB$ is a geodesic of length $s_{12}$.  The
longitude of $B$ relative to $A$ is $\lambda_{12}$; the latitudes of $A$
and $B$ are $\phi_1$ and $\phi_2$.  $EF\;H$ is the equator with $E$ also
lying on the extension of the geodesic $AB$; and $\alpha_0$, $\alpha_1$,
and $\alpha_2$ are the azimuths (in the forward direction) of the
geodesic at $E$, $A$, and $B$.}
\end{figure}

The shortest path between two points on the earth, customarily treated
as an ellipsoid of revolution, is called a {\it geodesic}.  Two geodesic
problems are usually considered: the {\it direct} problem of finding the
end point of a geodesic given its starting point, initial azimuth, and
length; and the {\it inverse} problem of finding the shortest path
between two given points.  Referring to Fig.~\ref{figtrig}, it can be
seen that each problem is equivalent to solving the geodesic triangle
$N\;AB$ given two sides and their included angle (the azimuth at the
first point, $\alpha_1$, in the case of the direct problem and the
longitude difference, $\lambda_{12}$, in the case of the inverse
problem).  The framework for solving these problems was laid down
by \citet{legendre06}, \citet{oriani06,oriani08,oriani10}, \citet{bessel25},
and \citet{helmert80}.  Based on these works, \citet{vincenty75a}
devised algorithms for solving the geodesic problems suitable for early
programmable desk calculators; these algorithms are in widespread use
today.  A good summary of Vincenty's algorithms and the earlier work in
the field is given by \citet[Chap.~1]{rapp93}.

The goal of this paper is to adapt the geodesic methods
of \citet{helmert80} and his predecessors to modern computers.  The
current work goes beyond Vincenty in three ways: (1)~The accuracy is
increased to match the standard precision of most computers.  This is a
relatively straightforward task of retaining sufficient terms in the
series expansions and can be achieved at little computational cost.
(2)~A solution of the inverse problem is given which converges for all
pairs of points.  (Vincenty's method fails to converge for nearly
antipodal points.)  (3)~Differential and integral properties of the
geodesics are computed.  The differential properties allow the behavior
of nearby geodesics to be determined, which enables the scales of
geodesic projections to be computed without resorting to numerical
differentiation; crucially, one of the differential quantities is also
used in the solution of the inverse problem.  The integral properties
provide a method for finding the area of a geodesic polygon, extending
the work of \citet{danielsen89}.

Section \ref{basic} reviews the classical solution of geodesic problem
by means of the auxiliary sphere and provides expansions of the
resulting integrals accurate to $O(f^6)$ (where $f$ is the flattening of
the ellipsoid).  These expansions can be inserted into the solution for
the direct geodesic problem presented by, for example, \citet{rapp93} to
provide accuracy to machine precision.  Section \ref{redlength} gives
the differential properties of geodesics reviewing the results
of \citet{helmert80} for the reduced length and geodesic scale and give
the key properties of these quantities and appropriate series expansions
to allow them to be calculated accurately.  Knowledge of the reduced
length enables the solution of the inverse problem by Newton's method
which is described in Sect.~\ref{inverse}.  Newton's method requires a
good starting guess and, in the case of nearly antipodal points, this is
provided by an approximate solution of the inverse problem
by \citet{helmert80}, as given in Sect.~\ref{antipodal}.  The
computation of area between a geodesic and the equator is formulated in
Sect.~\ref{area}, extending the work of \citet{danielsen89}.  Some
details of the implementation and present accuracy and timing data are
discussed in Sect.~\ref{implement}.  As an illustration of the use of
these algorithms, Sect.~\ref{gnomproj} gives an ellipsoidal gnomonic
projection in which geodesics are very nearly straight.  This provides a
convenient way of solving several geodesic problems.

For the purposes of this paper, it is useful to generalize the
definition of a geodesic.  The geodesic curvature, $\kappa$, of an
arbitrary curve at a point $P$ on a surface is defined as the curvature
of the projection of the curve onto a plane tangent to the surface at
$P$.  All shortest paths on a surface are {\it straight}, defined as
$\kappa = 0$ at every point on the path.  In the rest of this paper, I
use straightness as the defining property of geodesics; this allows
geodesic lines to be extended indefinitely (beyond the point at which
they cease to be shortest paths).

Several of the results reported here appeared earlier in a technical
report, \citet{karney11a}.

\section{Basic Equations and Direct Problem}\label{basic}

\begin{figure}[tb]
\begin{center}
\includegraphics[scale=0.75,angle=0]{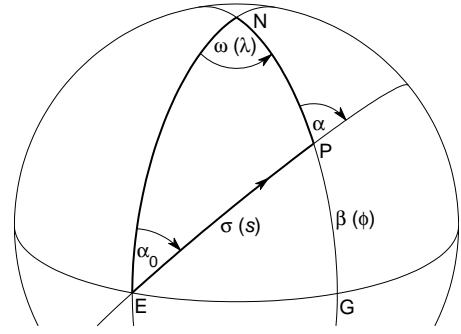}
\end{center}
\caption{\label{figtri}
The elementary ellipsoidal triangle $N\;EP$ mapped to the auxiliary
sphere.  $N\;E$ and $N\;PG$ are meridians; $EG$ is the equator; and $EP$
is the great circle (i.e., the geodesic).  The corresponding ellipsoidal
variables are shown in parentheses.  Here $P$ represents an arbitrary
point on the geodesic $EAB$ in Fig.~\ref{figtrig}.}
\end{figure}

I consider an ellipsoid of revolution with equatorial radius $a$, and
polar semi-axis $b$, flattening $f$, third flattening $n$, eccentricity
$e$, and second eccentricity $e'$ given by
\begin{align}
f &= (a - b) / a = 1 - \sqrt{1 - e^2},\label{feq}\displaybreak[0]\\
n &= (a - b)/(a + b) = f/(2-f),\label{neq}\displaybreak[0]\\
e^2 &= (a^2 - b^2)/a^2 = f(2-f),\label{e2eq}\displaybreak[0]\\
e'^2 &= (a^2 - b^2)/b^2 = e^2/(1-e^2).\label{e12eq}
\end{align}
As a consequence of the rotational symmetry of the ellipsoid, geodesics
obey a relation found by \citet{clairaut35}, namely
\begin{equation}
\sin\alpha_0 = \sin\alpha_1 \cos\beta_1
= \sin\alpha_2 \cos\beta_2,\label{napa0a}
\end{equation}
where $\beta$ is the reduced latitude (sometimes called the parametric
latitude), given by
\begin{equation}\label{redlat}
\tan\beta = (1-f)\tan\phi.
\end{equation}
The geodesic problems are most easily solved by using an {\it auxiliary
sphere} which allows an exact correspondence to be made between a
geodesic and a great circle on a sphere.  On the sphere, the latitude
$\phi$ is replaced by the reduced latitude $\beta$, and azimuths
$\alpha$ are preserved.  From Fig.~\ref{figtri}, it is clear that
Clairaut's equation, $\sin\alpha_0 = \sin\alpha \cos\beta$, is just the
sine rule applied to the sides $N\;E$ and $N\;P$ of the triangle $N\;EP$
and their opposite angles.  The third side, the spherical arc length
$\sigma$, and its opposite angle, the spherical longitude $\omega$, are
related to the equivalent quantities on the ellipsoid, the distance $s$
and longitude $\lambda$, by \citep[Eqs.~(1.28) and (1.170)]{rapp93}
\begin{align}
\frac sb &= \int_0^\sigma\sqrt{1 + k^2\sin^2\sigma'}\,\d\sigma'
= I_1(\sigma), \label{disteq}\displaybreak[0]\\
\lambda &= \omega - f\sin\alpha_0
\int_0^\sigma \frac{2-f}{1+(1-f)\sqrt{1 + k^2\sin^2\sigma'}}
\,\d\sigma'\notag\\
&= \omega - f\sin\alpha_0 \,I_3(\sigma), \label{lameq}
\end{align}
where
\begin{equation}\label{keq}
k = e'\cos\alpha_0.
\end{equation}
See also Eqs.~(5.4.9) and (5.8.8) of \citet{helmert80}.  The origin for
$s$, $\sigma$, $\lambda$, and $\omega$ is the point $E$, at which the
geodesic crosses the equator in the northward direction, with azimuth
$\alpha_0$.  The point $P$ can stand for either end of the geodesic $AB$
in Fig.~\ref{figtrig}, with the quantities $\beta$, $\alpha$, $\sigma$,
$\omega$, $s$, and $\lambda$ acquiring a subscript $1$ or $2$.  I also
define $s_{12} = s_2 - s_1$ as the length of $AB$, with $\lambda_{12}$,
$\sigma_{12}$, and $\omega_{12}$ defined similarly.  (In this paper,
$\alpha_2$ is the forward azimuth at $B$.  Several authors use the back
azimuth instead; this is given by $\alpha_2 \pm \pi$.)

Because Eqs.~(\ref{disteq}) and (\ref{lameq}) depend on $\alpha_0$, the
mapping between the ellipsoid and the auxiliary sphere is not a global
mapping of one surface to another; rather the auxiliary sphere should
merely be regarded as a useful mathematical technique for solving
geodesic problems.  Similarly, because the origin for $\lambda$ depends
on the geodesic, only longitude differences, e.g., $\lambda_{12}$,
should be used in converting between longitudes relative to the prime
meridian and $\lambda$.

In solving the spherical trigonometrical problems, the following
equations relating the sides and angles of $N\;EP$ are useful,
\begin{align}
\alpha_0 &= \ph(\abs{\cos\alpha + i \sin\alpha\sin\beta}
  + i \sin\alpha\cos\beta),\label{alp0eq}\displaybreak[0]\\
\sigma &= \ph(\cos\alpha \cos\beta + i \sin\beta),
\label{sigeq}\displaybreak[0]\\
\omega &= \ph(\cos\sigma + i \sin\alpha_0 \sin\sigma),
\label{omgeq}\displaybreak[0]\\
\beta &= \ph(\abs{\cos\alpha_0 \cos\sigma + i \sin\alpha_0}
  + i \cos\alpha_0 \sin\sigma),\label{beteq}\displaybreak[0]\\
\alpha &= \ph(\cos\alpha_0 \cos\sigma + i \sin\alpha_0),\label{alpeq}
\end{align}
where $i = \sqrt{-1}$ and $\ph(x + iy)$ is the phase of a
complex number \citep[\S\dlmf{1.9(i)}{1.9.i}]{dlmf10}, typically given
by the library function $\atantwo(y, x)$.  Equation
(\ref{alp0eq}) merely recasts Eq.~(\ref{napa0a}) in a form that allows
it to be evaluated accurately when $\alpha_0$ is close to $\frac12 \pi$.
The other relations are obtained by applying Napier's rules of circular
parts to $N\;EP$.

The distance integral, Eq.~(\ref{disteq}), can be expanded in a Fourier
series
\begin{equation}\label{i1expr}
I_1(\sigma) = A_1 \Bigl(\sigma +
\sum_{l = 1}^\infty C_{1l}\sin 2l\sigma\Bigr),
\end{equation}
with the coefficients determined by expanding the integral in a Taylor
series.  It is advantageous to follow \citet[\S5]{bessel25}
and \citet[Eq.~(5.5.1)]{helmert80} and use $\epsilon$, defined by
\begin{equation}
\epsilon =\frac{\sqrt{1+k^2}-1}{\sqrt{1+k^2}+1}\quad\text{or}
\quad k = \frac{2\sqrt\epsilon}{1-\epsilon},\label{epseqa}
\end{equation}
as the expansion parameter.  This leads to expansions with half as many
terms as the corresponding ones in $k^2$.  The expansion can be
conveniently carried out to arbitrary order by a computer algebra system
such as \citet{maxima} which yields
\begin{align}
A_1 &= (1 - \epsilon)^{-1} \bigl(1 + \tfrac1/4 \epsilon^2
+ \tfrac1/64 \epsilon^4 + \tfrac1/256 \epsilon^6 +\cdots\bigr),
\label{A1}\displaybreak[0]\\
C_{11} &= - \tfrac1/2 \epsilon + \tfrac3/16 \epsilon^3
- \tfrac1/32 \epsilon^5 + \cdots,\displaybreak[0]\notag\\
C_{12} &= - \tfrac1/16 \epsilon^2 + \tfrac1/32 \epsilon^4
- \tfrac9/2048 \epsilon^6 + \cdots,\displaybreak[0]\notag\\
C_{13} &= - \tfrac1/48 \epsilon^3 + \tfrac3/256 \epsilon^5
+ \cdots,\displaybreak[0]\notag\\
C_{14} &= - \tfrac5/512 \epsilon^4 + \tfrac3/512 \epsilon^6
+ \cdots,\displaybreak[0]\notag\\
C_{15} &= - \tfrac7/1280 \epsilon^5
+ \cdots,\displaybreak[0]\notag\\
C_{16} &= - \tfrac7/2048 \epsilon^6 + \cdots.
\label{C1}
\end{align}
This extends Eq.~(5.5.7) of \citet{helmert80} to higher order.  These
coefficients may be inserted into Eq.~(1.40) of \citet{rapp93} using
\begin{equation}\label{rapp1}
\mathcal B_j = \begin{cases}
A_1,& \text{for $j = 0$},\\
2A_1C_{1l},& \text{for $j = 2l$, with $l>0$},
\end{cases}
\end{equation}
where here, and subsequently in Eqs.~(\ref{rapp2}) and (\ref{rapp3}), a
script letter, e.g., $\mathcal B$, is used to stand for Rapp's
coefficients.

In the course of solving the direct geodesic problem (where $s_{12}$
is given), it is necessary to determine $\sigma$ given $s$.  Vincenty
solves for $\sigma$ iteratively.  However, it is simpler to follow
\citet[\S5.6]{helmert80} and substitute $s = bA_1\tau$ into
Eqs.~(\ref{disteq}) and (\ref{i1expr}), to obtain $\tau = \sigma
+ \sum_l C_{1l}\sin 2l\sigma$; this may be inverted, for example, using
Lagrange reversion, to give
\begin{equation}
\sigma = \tau + \sum_{l = 1}^\infty C'_{1l}\sin 2l\tau,\label{sigmaeq}
\end{equation}
where
\begin{align}
C'_{11} &= \tfrac1/2 \epsilon - \tfrac9/32 \epsilon^3
+ \tfrac205/1536 \epsilon^5 + \cdots,\displaybreak[0]\notag\\
C'_{12} &= \tfrac5/16 \epsilon^2 - \tfrac37/96 \epsilon^4
+ \tfrac1335/4096 \epsilon^6 + \cdots,\displaybreak[0]\notag\\
C'_{13} &= \tfrac29/96 \epsilon^3 - \tfrac75/128 \epsilon^5
+ \cdots,\displaybreak[0]\notag\\
C'_{14} &= \tfrac539/1536 \epsilon^4 - \tfrac2391/2560 \epsilon^6
+ \cdots,\displaybreak[0]\notag\\
C'_{15} &= \tfrac3467/7680 \epsilon^5
+ \cdots,\displaybreak[0]\notag\\
C'_{16} &= \tfrac38081/61440 \epsilon^6 + \cdots.
\label{C11}
\end{align}
This extends Eq.~(5.6.8) of \citet{helmert80} to higher order.  These
coefficients may be used in Eq.~(1.142) of \citet{rapp93} using
\begin{equation}\label{rapp2}
\mathcal D_j = 2C'_{1l},\quad\text{for $j = 2l$, with $l>0$}.
\end{equation}

Similarly, the integral appearing in the longitude equation,
Eq.~(\ref{lameq}), can be written as a Fourier series
\begin{equation}
I_3(\sigma) = A_3 \Bigl(\sigma +
\sum_{l = 1}^\infty C_{3l}\sin 2l\sigma\Bigr).\label{i3expr}
\end{equation}
Following \citet{helmert80}, I expand jointly in $n$ and $\epsilon$,
both of which are $O(f)$, to give
\begin{align}
A_3 &= 1 - \bigl(\tfrac1/2 - \tfrac1/2 n\bigr) \epsilon
- \bigl(\tfrac1/4 + \tfrac1/8 n - \tfrac3/8
n^2\bigr) \epsilon^2 \notag\\&\qquad{} - \bigl(\tfrac1/16 + \tfrac3/16 n
+ \tfrac1/16 n^2\bigr) \epsilon^3 - \bigl(\tfrac3/64 + \tfrac1/32
n\bigr) \epsilon^4 \notag\\&\qquad{} - \tfrac3/128 \epsilon^5+ \cdots,
\label{A3}\displaybreak[0]\\
C_{31} &= \bigl(\tfrac1/4 - \tfrac1/4 n\bigr) \epsilon
+ \bigl(\tfrac1/8 - \tfrac1/8 n^2\bigr) \epsilon^2 + \bigl(\tfrac3/64
+ \tfrac3/64 n - \tfrac1/64 n^2\bigr) \epsilon^3 \notag\\&\quad{}
+ \bigl(\tfrac5/128 + \tfrac1/64 n\bigr) \epsilon^4
+ \tfrac3/128 \epsilon^5 + \cdots,\displaybreak[0]\notag\\
C_{32} &= \bigl(\tfrac1/16 - \tfrac3/32 n + \tfrac1/32
n^2\bigr) \epsilon^2 + \bigl(\tfrac3/64 - \tfrac1/32 n - \tfrac3/64
n^2\bigr) \epsilon^3 \notag\\&\quad{} + \bigl(\tfrac3/128 + \tfrac1/128
n\bigr) \epsilon^4 + \tfrac5/256 \epsilon^5
+ \cdots, \displaybreak[0]\notag\\
C_{33} &= \bigl(\tfrac5/192 - \tfrac3/64 n + \tfrac5/192
n^2\bigr) \epsilon^3 + \bigl(\tfrac3/128 - \tfrac5/192
n\bigr) \epsilon^4 \notag\\&\quad{} + \tfrac7/512 \epsilon^5
+ \cdots,\displaybreak[0]\notag\\
C_{34} &= \bigl(\tfrac7/512 - \tfrac7/256 n\bigr) \epsilon^4
+ \tfrac7/512 \epsilon^5 + \cdots,\displaybreak[0]\notag\\
C_{35} &= \tfrac21/2560 \epsilon^5 + \cdots.
\label{C3}
\end{align}
This extends Eq.~(5.8.14) of \citet{helmert80} to higher order.  These
coefficients may be inserted into Eq.~(1.56) of \citet{rapp93} using
\begin{equation}\label{rapp3}
\mathcal A_j = \begin{cases}
A_3,& \text{for $j = 0$},\\
2A_3C_{3l},& \text{for $j = 2l$, with
$l>0$}.
\end{cases}
\end{equation}

\begin{table}[tb]
\caption{\label{wgs84}
The parameters for the WGS84 ellipsoid used in the examples.  The column
labeled ``Eq.''\ lists the equations used to compute the corresponding
quantities.}
% printparams();
\begin{center}
\begin{tabular}{@{\extracolsep{0.5\xx}}>{$}c<{$} >{$}l<{$} >{}c<{}}
\hline\hline\noalign{\smallskip}
\text{Qty.} & \text{Value} & Eq.
\\\noalign{\smallskip}\hline\noalign{\smallskip}
a& 6\,378\,137 \,\mathrm m&given \\
f&1/ 298.257\,223\,563 &given \\
b& 6\,356\,752.314\,245\,\mathrm m & (\ref{feq}) \\
c& 6\,371\,007.180\,918\,\mathrm m & (\ref{authalic}) \\
n& 0.001\,679\,220\,386\,383\,70 & (\ref{neq}) \\
e^2& 0.006\,694\,379\,990\,141\,32 & (\ref{e2eq}) \\
e'^2& 0.006\,739\,496\,742\,276\,43 & (\ref{e12eq}) \\
\noalign{\smallskip}\hline\hline
\end{tabular}
\end{center}
\end{table}%
\begin{table}[tb]
\caption{\label{directex}
A sample direct calculation specified by $\phi_1 = 40^\circ$, $\alpha_1 =
30^\circ$, and $s_{12} = 10\,000\,\mathrm{km}$.  For equatorial
geodesics ($\phi_1 = 0$ and $\alpha_1 = \frac12 \pi$), Eq.~(\ref{sigeq})
is indeterminate; in this case, take $\sigma_1 = 0$.}
% directa(40, 30, 10b6);
\begin{center}
\begin{tabular}{@{\extracolsep{0.5\xx}}>{$}c<{$} >{$}l<{$} >{}c<{}}
\hline\hline\noalign{\smallskip}
\text{Qty.} & \text{Value} & Eq.
\\\noalign{\smallskip}\hline\noalign{\smallskip}
\phi_1& 40 ^\circ&given\\
\alpha_1& 30 ^\circ&given\\
s_{12}& 10\,000\,000 \,\mathrm m&given\\
\hline\multicolumn{3}{l}{Solve triangle $N\;EA$}\\
\beta_1 & 39.905\,277\,146\,01^\circ &(\ref{redlat})\\
\alpha_0 & 22.553\,940\,202\,62^\circ &(\ref{alp0eq})\\
\sigma_1 & 43.999\,153\,645\,00^\circ &(\ref{sigeq})\\
\omega_1 & 20.323\,718\,278\,37^\circ &(\ref{omgeq})\\
\hline\multicolumn{3}{l}{Determine $\sigma_2$}\\
k^2 & 0.005\,748\,029\,628\,57 &(\ref{keq})\\
\epsilon & 0.001\,432\,892\,204\,16 &(\ref{epseqa})\\
A_1 & 1.001\,435\,462\,362\,07 &(\ref{A1})\\
I_1(\sigma_1) & 0.768\,315\,388\,864\,12 &(\ref{i1expr})\\
s_1 &\hphantom{0} 4\,883\,990.626\,232\,\mathrm m &(\ref{disteq})\\
s_2 & 14\,883\,990.626\,232\,\mathrm m &$s_1 + s_{12}$\\
\tau_2 & 133.962\,660\,502\,08^\circ &$s_2/(bA_1)$\\
\sigma_2 & 133.921\,640\,830\,38^\circ &(\ref{sigmaeq})\\
\hline\multicolumn{3}{l}{Solve triangle $N\;EB$}\\
\alpha_2 & 149.090\,169\,318\,07^\circ &(\ref{alpeq})\\
\beta_2 &\hphantom{0} 41.697\,718\,092\,50^\circ &(\ref{beteq})\\
\omega_2 & 158.284\,121\,471\,12^\circ &(\ref{omgeq})\\
\hline\multicolumn{3}{l}{Determine $\lambda_{12}$}\\
A_3 & 0.999\,284\,243\,06 &(\ref{A3})\\
I_3(\sigma_1) & 0.767\,737\,860\,69 &(\ref{i3expr})\\
I_3(\sigma_2) & 2.335\,343\,221\,70 &(\ref{i3expr})\\
\lambda_1 &\hphantom{0} 20.267\,150\,380\,16^\circ &(\ref{lameq})\\
\lambda_2 & 158.112\,050\,423\,93^\circ &(\ref{lameq})\\
\lambda_{12} & 137.844\,900\,043\,77^\circ &$\lambda_2 - \lambda_1$\\
\hline\multicolumn{3}{l}{Solution}\\
\phi_2 &\hphantom{0} 41.793\,310\,205\,06^\circ &(\ref{redlat})\\
\lambda_{12} & 137.844\,900\,043\,77^\circ \\
\alpha_2 & 149.090\,169\,318\,07^\circ \\
\noalign{\smallskip}\hline\hline
\end{tabular}
\end{center}
\end{table}%
The equations given in this section allow the direct geodesic problem to
be solved.  Given $\phi_1$ (and hence $\beta_1$) and $\alpha_1$ solve
the spherical triangle $N\;EA$ to give $\alpha_0$, $\sigma_1$, and
$\omega_1$ using Eqs.~(\ref{alp0eq}), (\ref{sigeq}), and (\ref{omgeq}).
Find $s_1$ and $\lambda_1$ from Eqs.~(\ref{disteq}) and (\ref{lameq})
together with Eqs.~(\ref{i1expr}) and (\ref{i3expr}).  (Recall that the
origin for $\lambda$ is $E$ in Fig.~\ref{figtrig}.)  Determine $s_2 =
s_1 + s_{12}$ and hence $\sigma_2$ using Eq.~(\ref{sigmaeq}).  Now solve
the spherical triangle $N\;EB$ to give $\alpha_2$, $\beta_2$ (and hence
$\phi_2$), and $\omega_2$, using Eqs.~(\ref{alpeq}), (\ref{beteq}), and
(\ref{omgeq}).  Finally, determine $\lambda_2$ (and $\lambda_{12}$) from
Eqs.~(\ref{lameq}) and (\ref{i3expr}).  A numerical example of the
solution of the direct problem is given in Table \ref{directex} using
the parameters of Table \ref{wgs84}.

\section{Differential Quantities}\label{redlength}

Before turning to the inverse problem, I present Gauss' solution for the
differential behavior of geodesics.  One differential quantity, the
reduced length $m_{12}$, is needed in the solution of the inverse problem
by Newton's method (Sect.~\ref{inverse}) and an expression for this
quantity is given at the end of this section.  However, because this and
other differential quantities aid in the solution of many geodesic
problems, I also discuss their derivation and present some of their
properties.

Consider a reference geodesic parametrized by distance $s$ and a
nearby geodesic separated from the reference by infinitesimal distance $t(s)$.
\citet{gauss27} showed that
$t(s)$ satisfies the differential equation
\begin{equation}\label{reducedeq}
\frac{\d^2t(s)}{\d s^2} + K(s)\,t(s) = 0,
\end{equation}
where $K(s)$ is the Gaussian curvature of the surface.  As a second
order, linear, homogeneous differential equation, its solution
can be written as
\[
t(s) = A t_A(s) + B t_B(s),
\]
where $A$ and $B$ are (infinitesimal)
constants and $t_A$ and $t_B$ are independent
solutions.  When considering the geodesic segment spanning
$s_1 \le s \le s_2$, it is convenient to specify
\begin{align*}
t_A(s_1) &= 0, \quad \left. \frac{\d t_A(s)}{\d s} \right|_{s = s_1} = 1,
\displaybreak[0]\\
t_B(s_1) &= 1, \quad \left. \frac{\d t_B(s)}{\d s} \right|_{s = s_1} = 0,
\end{align*}
and to write
\[
m_{12} = t_A(s_2), \quad M_{12} = t_B(s_2).
\]
\begin{figure}[tb]
\begin{center}
\includegraphics[scale=0.75,angle=0]{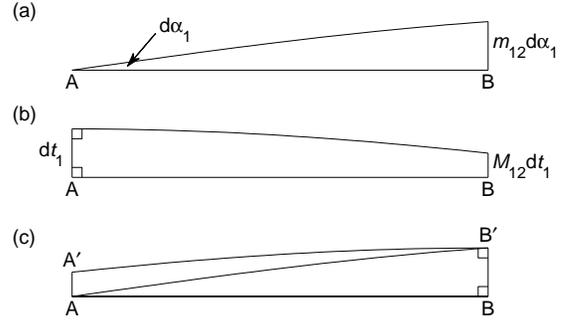}
\end{center}
\caption{\label{figred}
The definitions of $m_{12}$ and $M_{12}$ are illustrated in~(a) and~(b).
A geometric proof of Eq.~(\ref{m-deriv}) is shown in~(c); here $AB$ and
$A'B'$ are parallel at $B$ and $B'$, $BAB' = \d\alpha_1$, $BB' =
m_{12}\,\d\alpha_1$, $AA' = M_{21}m_{12}\,\d\alpha_1$, and finally
$AB'A' = M_{21}\,\d\alpha_1$, from which Eq.~(\ref{m-deriv}) follows.}
\end{figure}%
The quantity $m_{12}$ is the {\it reduced length} of the geodesic
\citep{christoffel68}.  Consider two geodesics which cross at $s = s_1$
at a small angle $\d\alpha_1$, Fig~\ref{figred}(a); at $s = s_2$, they
will be separated by a distance $m_{12}\,\d\alpha_1$. Similarly I call
$M_{12}$ the {\it geodesic scale}.  Consider two geodesics which are
parallel at $s = s_1$ and separated by a small distance $\d t_1$,
Fig~\ref{figred}(b); at $s = s_2$, they will be separated by a
distance $M_{12}\,\d t_1$.

Several relations between $m_{12}$ and $M_{12}$ follow from the defining
equation, Eq.~(\ref{reducedeq}).  The reduced length obeys a reciprocity
relation \citep[\S9]{christoffel68}, $m_{21} + m_{12} = 0$; the
Wronskian is given by
\begin{equation}\label{wronski}
W(M_{12}, m_{12})(s_2) =
M_{12}\frac{\d m_{12}}{\d s_2}-m_{12}\frac{\d M_{12}}{\d s_2} = 1;
\end{equation}
and the derivatives are
\begin{align}
\frac{\d m_{12}}{\d s_2} &= M_{21}, \label{m-deriv}
\displaybreak[0]\\
\frac{\d M_{12}}{\d s_2} &= -\frac{1 - M_{12}M_{21}}{m_{12}}. \label{M-deriv}
\end{align}
The constancy of the Wronskian follows by noting that its derivative
with respect to $s_2$ vanishes; its value is found by evaluating it at
$s_2 = s_1$.
A geometric proof of Eq.~(\ref{m-deriv}) is given in Fig~\ref{figred}(c)
and Eq.~(\ref{M-deriv}) then follows from Eq.~(\ref{wronski}).  With
knowledge of the derivatives, addition rules for $m_{12}$ and $M_{12}$
are easily found,
\begin{align}
m_{13} &= m_{12} M_{23} + m_{23} M_{21},\displaybreak[0]\\
M_{13} &= M_{12} M_{23} - (1 - M_{12} M_{21}) \frac{m_{23}}{m_{12}},
\displaybreak[0]\\
M_{31} &= M_{32} M_{21} - (1 - M_{23} M_{32}) \frac{m_{12}}{m_{23}},
\end{align}
where points 1, 2, and 3 all lie on the same geodesic.

Geodesics allow concepts from plane geometry to be generalized to apply
to a curved surface.  In particular, a geodesic circle may be defined as
the curve which is a constant geodesic distance from a fixed point.
Similarly, a geodesic parallel to a reference curve is the curve which
is a constant geodesic distance from that curve.  (Thus a circle is a
special case of a parallel obtained in the limit when the reference
curve degenerates to a point.)  Parallels occur naturally when
considering, for example, the ``12-mile limit'' for territorial waters
which is the boundary of points lying within 12 nautical miles of a
coastal state.

The geodesic curvature of a parallel can be expressed in terms of
$m_{12}$ and $M_{12}$.  Let point 1 be an arbitrary point on the
reference curve with geodesic curvature $\kappa_1$.  Point 2 is the
corresponding point on the parallel, a fixed distance $s_{12}$ away.
The geodesic curvature of the parallel at that point is found from
Eqs.~(\ref{m-deriv}) and (\ref{M-deriv}),
\begin{equation}
\kappa_2 =
\frac{M_{21} \kappa_1 - (1 - M_{12} M_{21})/m_{12}}
{m_{12} \kappa_1 + M_{12}}.\label{kappa2}
\end{equation}
The curvature of a circle is given by the limit $\kappa_1 \rightarrow
\infty$,
\begin{equation}
    \kappa_2 = M_{21} / m_{12}.
\end{equation}
If the reference curve is a geodesic ($\kappa_1 \rightarrow 0$), then
the curvature of its parallel is
\begin{equation}
    \kappa_2 = - (1 - M_{12} M_{21}) / (M_{12} m_{12}).
\end{equation}
If the reference curve is indented, then the parallel intersects itself
at a sufficiently large distance from the reference curve.  This begins
to happen when $\kappa_2 \rightarrow \infty$ in Eq.~(\ref{kappa2}).

The results above apply to general surfaces.  For a geodesic on an
ellipsoid of revolution, the Gaussian curvature of the surface is given
by
\begin{equation}
K = \frac{(1 - e^2 \sin^2\phi)^2}{ b^2 }
= \frac1 {b^2 (1 +  k^2 \sin^2\sigma)^2}. \label{curvature}
\end{equation}
\citet[Eq.~(6.5.1)]{helmert80} solves Eq.~(\ref{reducedeq}) in this
case to give
\begin{align}
m_{12}/b &= \sqrta{1 + k^2\sin^2\sigma_2}\, \cos\sigma_1 \sin\sigma_2
\notag\\&\quad - \sqrta{1 + k^2\sin^2\sigma_1}\, \sin\sigma_1 \cos\sigma_2
\notag\\ &\quad - \cos\sigma_1 \cos\sigma_2
\bigl(J(\sigma_2) - J(\sigma_1)\bigr),
\label{meq}\displaybreak[0]\\
M_{12} &= \cos\sigma_1 \cos\sigma_2
+ \frac{\sqrt{1 + k^2\sin^2\sigma_2}}{\sqrt{1 + k^2\sin^2\sigma_1}}
\sin\sigma_1 \sin\sigma_2
\notag\\&\quad - \frac{\sin\sigma_1 \cos\sigma_2
\bigl(J(\sigma_2) - J(\sigma_1)\bigr)}
{\sqrt{1 + k^2\sin^2\sigma_1}},\label{Meq}
\end{align}
where
\begin{align}
\label{jeq}
J(\sigma) &=
\int_0^\sigma \frac{k^2\sin^2\sigma'}{\sqrt{1 + k^2\sin^2\sigma'}}\,\d\sigma'
\notag\\&= \frac sb -
\int_0^\sigma \frac 1{\sqrt{1 + k^2\sin^2\sigma'}}\,\d\sigma'
\notag\\&= I_1(\sigma) - I_2(\sigma).
\end{align}
Equation (\ref{Meq}) may be obtained from Eq.~(6.9.7)
of \citet{helmert80}, which gives $\d m_{12}/\d s_2$; $M_{12}$ may then
be found from Eq.~(\ref{m-deriv}) with an interchange of indices.  In
the spherical limit, $f \rightarrow 0$, Eqs.~(\ref{meq}) and (\ref{Meq})
reduce to
\begin{align*}
m_{12} &= a \sin\sigma_{12} = a \sin(s_{12}/a), \\
M_{12} &= \cos\sigma_{12} = \cos(s_{12}/a).
\end{align*}

The integral $I_2(\sigma)$ in Eq.~(\ref{jeq}) may be expanded in a
Fourier series in similar fashion to $I_1(\sigma)$, Eq.~(\ref{i1expr}),
\begin{equation}
I_2(\sigma) = A_2 \Bigl(\sigma +
\sum_{l = 1}^\infty C_{2l}\sin 2l\sigma\Bigr),\label{i2expr}
\end{equation}
where
\begin{align}
A_2 &= (1 - \epsilon) \bigl(1 + \tfrac1/4 \epsilon^2
+ \tfrac9/64 \epsilon^4 + \tfrac25/256 \epsilon^6 + \cdots\bigr),
\label{A2}\displaybreak[0]\\
C_{21} &= \tfrac1/2 \epsilon + \tfrac1/16 \epsilon^3
+ \tfrac1/32 \epsilon^5 + \cdots,\displaybreak[0]\notag\\
C_{22} &= \tfrac3/16 \epsilon^2 + \tfrac1/32 \epsilon^4
+ \tfrac35/2048 \epsilon^6 + \cdots,\displaybreak[0]\notag\\ C_{23}
&= \tfrac5/48 \epsilon^3 + \tfrac5/256 \epsilon^5
+ \cdots,\displaybreak[0]\notag\\
C_{24} &= \tfrac35/512 \epsilon^4 + \tfrac7/512 \epsilon^6
+ \cdots,\displaybreak[0]\notag\\
C_{25} &= \tfrac63/1280 \epsilon^5
+ \cdots,\displaybreak[0]\notag\\ C_{26}
&= \tfrac77/2048 \epsilon^6.
\label{C2}
\end{align}

\section{Inverse Problem}\label{inverse}

The inverse problem is intrinsically more complicated than the direct
problem because the given included angle, $\lambda_{12}$ in
Fig.~\ref{figtrig}, is related to the corresponding angle on the
auxiliary sphere $\omega_{12}$ via an unknown equatorial azimuth
$\alpha_0$.  Thus, the inverse problem inevitably becomes a root-finding
exercise.

I tackle this problem as follows.  Assume that $\alpha_1$ is known.
Solve the {\it hybrid} geodesic problem: given $\phi_1$, $\phi_2$, and
$\alpha_1$, find $\lambda_{12}$ corresponding to the first intersection
of the geodesic with the circle of latitude $\phi_2$.  The resulting
$\lambda_{12}$ differs, in general, from the given $\lambda_{12}$; so
adjust $\alpha_1$ using Newton's method until the correct $\lambda_{12}$
is obtained.

I begin by putting the points in a canonical configuration,
\begin{equation}\label{canon}
\phi_1 \le 0,\quad \phi_1 \le \phi_2 \le -\phi_1,\quad
 0\le \lambda_{12} \le \pi.
\end{equation}
This may be accomplished swapping the end points and the signs of the
coordinates if necessary, and the solution may similarly be transformed
to apply to the original points.  All geodesics with $\alpha_1 \in
[0,\pi]$ intersect latitude $\phi_2$ with $\lambda_{12} \in [0,\pi]$.
Furthermore, the search for solutions can be restricted to $\alpha_2 \in
[0, \frac12 \pi]$, because this corresponds to the first intersection
with latitude $\phi_2$.

Meridional ($\lambda_{12} = 0$ or $\pi$) and equatorial ($\phi_1
= \phi_2 = 0$, with $\lambda_{12} \le (1-f)\pi$) geodesics are treated
as special cases, since the azimuth is then known: $\alpha_1
= \lambda_{12}$ and $\alpha_1 = \frac12 \pi$ respectively.  The general
case is solved by Newton's method as outlined above.

\begin{figure}[tb]
\begin{center}
\includegraphics[scale=0.75,angle=0]{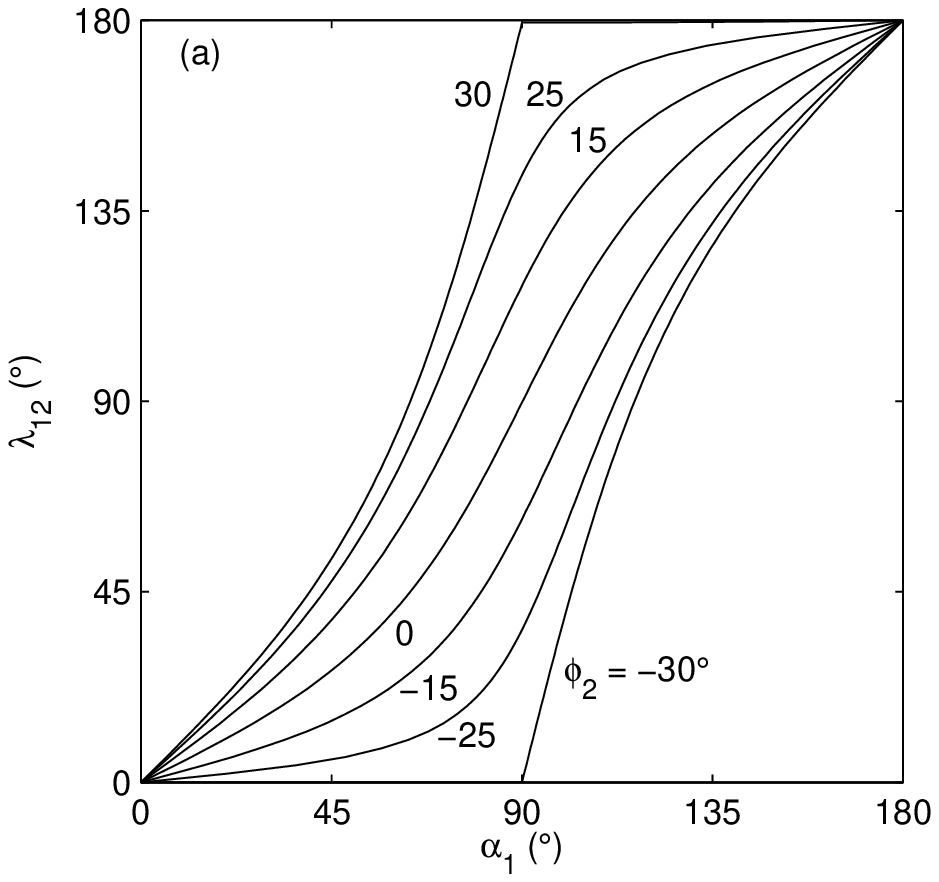}\\[5pt]
\includegraphics[scale=0.75,angle=0]{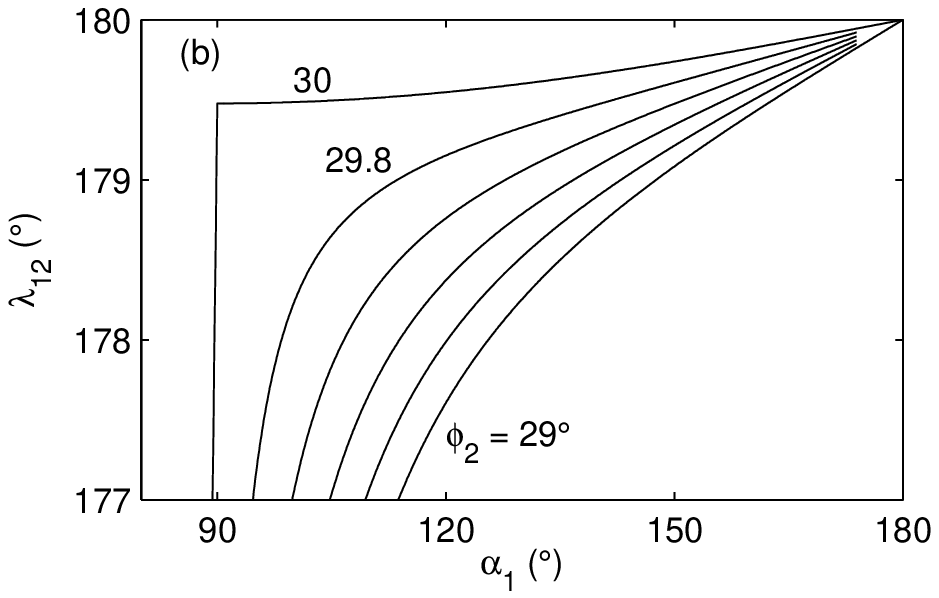}
\end{center}
\caption{\label{figalp1}
The variation of $\lambda_{12}$ as a function of $\alpha_1$ for $\phi_1
= -30^\circ$, various $\phi_2$, and the WGS84 ellipsoid.  Part~(a) shows
$\lambda_{12}$ for $\phi_2 = 0^\circ$, $\pm15^\circ$, $\pm25^\circ$, and
$\pm30^\circ$.  For $\abs{\phi_2} < \abs{\phi_1}$, the curves are
strictly increasing, while for $\phi_2 = \pm\phi_1$, the curves are
non-decreasing with discontinuities in the slopes at $\alpha_1 =
90^\circ$.  An enlargement of the top right corner of~(a) is shown
in~(b) with $\phi_2 \in [29^\circ, 30^\circ]$ at intervals of
$0.2^\circ$.}
\end{figure}%
The solution of the hybrid geodesic problem is straightforward.  Find
$\beta_1$ and $\beta_2$ from Eq.~(\ref{redlat}), solve for $\alpha_0$
and $\alpha_2$ from Eq.~(\ref{napa0a}), taking $\cos\alpha_0 > 0$ and
$\cos\alpha_2 \ge 0$.  In order to compute $\alpha_2$ accurately, use
\begin{equation}\label{alp2-eq}
\cos\alpha_2 = \frac{+\sqrt{\cos^2\alpha_1 \cos^2\beta_1 +
 (\cos^2\beta_2 - \cos^2\beta_1)}}{\cos\beta_2},
\end{equation}
in addition to Eq.~(\ref{napa0a}).
Compute $\sigma_1$, $\omega_1$, $\sigma_2$, and
$\omega_2$ using Eqs.~(\ref{sigeq}) and (\ref{omgeq}).
Finally, determine $\lambda_{12}$
(and, once convergence is achieved, $s_{12}$) as in the solution to the
direct problem.  The behavior of $\lambda_{12}$ as a function of
$\alpha_1$ is shown in Fig.~\ref{figalp1}.

\begin{figure}[tb]
\begin{center}
\includegraphics[scale=0.75,angle=0]{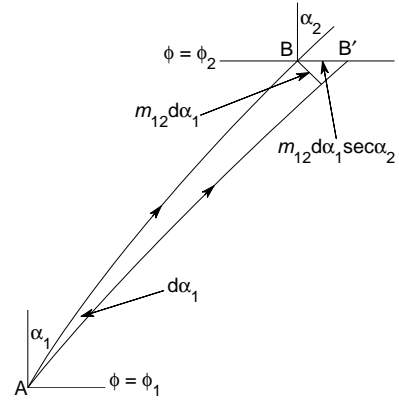}
\end{center}
\caption{\label{derivfig}
Finding $\d\lambda_{12}/\d\alpha_1$ with $\phi_1$ and $\phi_2$ held
fixed.}
\end{figure}

To apply Newton's method, an expression for $\d\lambda_{12}/\d\alpha_1$
is needed.  Consider a geodesic with initial azimuth $\alpha_1$.  If the
azimuth is increased to $\alpha_1 + \d\alpha_1$ with its length held
fixed, then the other end of the geodesic moves by $m_{12}\,\d\alpha_1$
in a direction $\frac12 \pi + \alpha_2$.  If the geodesic is extended to
intersect the parallel $\phi_2$ once more, the point of intersection
moves by $m_{12}\,\d\alpha_1/\cos\alpha_2$; see Fig.~\ref{derivfig}.
The radius of this parallel is $a\cos\beta_2$; thus the rate of change
of the longitude difference is
\begin{equation}\label{lambda-deriv}
\frac{\d\lambda_{12}}{\d\alpha_1}
= \frac{m_{12}}a
\frac1{\cos\alpha_2 \cos\beta_2}.
\end{equation}
This equation can also be obtained from Eq.~(6.9.8b)
of \citet{helmert80}.  Equation (\ref{lambda-deriv}) becomes
indeterminate when $\beta_2 = \pm\beta_1$ and $\alpha_1 = \frac12 \pi$,
because $m_{12}$ and $\cos\alpha_2$ both vanish. In this case, it is
necessary to let $\alpha_1 = \frac12 \pi + \delta$ and to take the
limit $\delta \rightarrow \pm0$, which gives
\begin{equation}\label{lambda-deriv-0}
\frac{\d\lambda_{12}}{\d\alpha_1}
= -\frac{\sqrt{1-e^2 \cos^2\beta_1}}{\sin\beta_1}
\bigl(1 \mp \sign(\cos\alpha_1) \bigr),
\end{equation}
where $\sign(\cos\alpha_1) = -\sign(\delta)$.  A numerical example of
solving the inverse geodesic problem by this method is given at the end
of the next section.

\citet{vincenty75a}, who uses the iterative method
of \citet[\S5.13]{helmert80} to solve the inverse problem, was aware of
its failure to converge for nearly antipodal points.  In an unpublished
report \citep{vincenty75b}, he gives a modification of his method which
deals with this case.  Unfortunately, this sometimes requires many
thousands of iterations to converge, whereas Newton's method as
described here only requires a few iterations.

\section{Starting point for Newton's method}\label{antipodal}

\begin{table}[tb]
\caption{\label{sphereinvex}
First sample inverse calculation specified by $\phi_1 =
-30.123\,45^\circ$, $\phi_2 = -30.123\,44^\circ$, and $\lambda_{12} =
0.000\,05^\circ$.  Because the points are not nearly antipodal, an
initial guess for $\alpha_1$ is found assuming $\omega_{12}
= \lambda_{12}/\bar w$.  However, in this case, the line is short enough
that the error in $\omega_{12}$ is negligible at the precision given and
the solution of the inverse problem is completed by using $s_{12} =
a \bar w \sigma_{12}$.  More generally, the value of $\alpha_1$ would be
refined using Newton's method.}
% inversestarta(-30.12345b0, -30.12344b0, 0.00005b0);
\begin{center}
\begin{tabular}{@{\extracolsep{0.5\xx}}>{$}c<{$} >{$}l<{$} >{}c<{}}
\hline\hline\noalign{\smallskip}
\text{Qty.} & \hphantom{-}\text{Value} & Eq.
\\\noalign{\smallskip}\hline\noalign{\smallskip}
\phi_1& -30.123\,45 ^\circ& given \\
\phi_2& -30.123\,44 ^\circ& given \\
\mbox{}\quad\lambda_{12}\quad\mbox{}&\hphantom{-0} 0.000\,05 ^\circ& given \\
\hline\multicolumn{3}{l}{Determine $\omega_{12}$}\\
\beta_1 & -30.039\,990\,838\,21^\circ &(\ref{redlat})\\
\beta_2 & -30.039\,980\,854\,91^\circ &(\ref{redlat})\\
\bar w&\hphantom{-0} 0.997\,488\,477\,44 &(\ref{wbar})\\
\omega_{12} &\hphantom{-0} 0.000\,050\,125\,89^\circ &$\lambda_{12}/\bar w$\\
\sigma_{12} &\hphantom{-0} 0.000\,044\,526\,41^\circ &(\ref{sigma-spher})\\
\hline\multicolumn{3}{l}{Solution}\\
\alpha_1 &\hphantom{-} 77.043\,533\,542\,37^\circ &(\ref{alpha1-spher})\\
\alpha_2 &\hphantom{-} 77.043\,508\,449\,13^\circ &(\ref{alpha2-spher})\\
s_{12} &\hphantom{-0} 4.944\,208\,\mathrm m &$a\bar w\sigma_{12}$\\
\noalign{\smallskip}\hline\hline
\end{tabular}
\end{center}
\end{table}%
To complete the solution of the inverse problem a good starting guess
for $\alpha_1$ is needed.  In most cases, this is provided by assuming
that $\omega_{12} = \lambda_{12}/\bar w$, where
\begin{equation}\label{wbar}
\bar w = \sqrt{ 1 - e^2 \bigl((\cos\beta_1 + \cos\beta_2)/2\bigr)^2}
\end{equation}
and solving for the great circle on the
auxiliary sphere, using \citep{vincenty75a}
\begin{align}
z_1 &=\cos\beta_1\sin\beta_2 - \sin\beta_1\cos\beta_2 \cos\omega_{12}
\notag\\&\qquad
+i\cos\beta_2\sin\omega_{12},\displaybreak[0]\notag\\
z_2 &= -\sin\beta_1\cos\beta_2 + \cos\beta_1\sin\beta_2 \cos\omega_{12}
\notag\\&\qquad
+i\cos\beta_1\sin\omega_{12},\displaybreak[0]\notag\\
\alpha_1 &= \ph z_1, \label{alpha1-spher}\displaybreak[0]\\
\alpha_2 &= \ph z_2, \label{alpha2-spher}\displaybreak[0]\\
\sigma_{12} &=
\ph(\sin\beta_1\sin\beta_2 + \cos\beta_1\cos\beta_2 \cos\omega_{12}
+ i \abs{z_1}). \label{sigma-spher}
\end{align}
An example of the solution of the inverse problem by this method is
given in Table \ref{sphereinvex}.

This procedure is inadequate for nearly antipodal points because both
the real and imaginary components of $z_1$ are small and $\alpha_1$
depends very sensitively on $\omega_{12}$.  In the corresponding
situation on the sphere, it is possible to determine $\alpha_1$ by
noting that all great circles emanating from $A$ meet at $O$, the point
antipodal to $A$.  Thus $\alpha_1$ may be determined as the supplement
of the azimuth of the great circle $BO$ at $O$; in addition, because $B$
and $O$ are close, it is possible to approximate the sphere, locally, as
a plane.

The situation for an ellipsoid is slightly different because the
geodesics emanating from $A$, instead of meeting at a point, form an
envelope, centered at $O$, in the shape of an {\it astroid} whose extent
is $O(f)$ \citep[Eqs.~(16)--(17)]{jacobi91}.  The position at which a
particular geodesic touches this envelope is given by the condition
$m_{12} = 0$.  However elementary methods can be used to determine the
envelope.  Consider a geodesic leaving $A$ (with $\beta_1 \le 0$) with
azimuth $\alpha_1 \in [\tfrac1/2 \pi, \pi]$.  This first intersects the
circle of opposite latitude, $\beta_2 = -\beta_1$, with $\sigma_{12}
= \omega_{12} = \pi$ and $\alpha_2 = \pi - \alpha_1$.  Equation
~(\ref{lameq}) then gives
\begin{equation}\label{astroid-lat}
\lambda_{12} = \pi - f \pi \cos\beta_1 \sin\alpha_1 + O(f^2).
\end{equation}
Define a plane coordinate system $(x,y)$ centered on the antipodal point
where $\Delta = fa \pi \cos^2\beta_1$ is the unit of length, i.e.,
\begin{equation}\label{xyeq}
\lambda_{12} = \pi + \frac {\Delta}{a\cos\beta_1}x, \quad
\beta_2 = -\beta_1 + \frac {\Delta}a y.
\end{equation}
In this coordinate system, Eq.~(\ref{astroid-lat}) corresponds to the
point $x = -\sin\alpha_1$, $y = 0$ and the slope of the geodesic is
$-\cot\alpha_1$.  Thus, in the neighborhood of the antipodal point, the
geodesic may be approximated by
\begin{equation}\label{anti-line}
\frac x {\sin\alpha_1} + \frac y {\cos\alpha_1} + 1 = 0,
\end{equation}
where terms of order $f^2$ have been neglected.  Allowing $\alpha_1$ to
vary, Eq.~(\ref{anti-line}) defines a family of lines approximating the
geodesics emanating from $A$.  Differentiating this equation with
respect to $\alpha_1$ and solving the resulting pair of equations for
$x$ and $y$ gives the parametric equations for the astroid, $x =
- \sin^3\alpha_1$, and $y = - \cos^3\alpha_1$.  Note that, for the
ordering of points given by Eq.~(\ref{canon}), $x \le 0$ and $y \le 0$.

\begin{figure}[tb]
\begin{center}
\includegraphics[scale=0.75,angle=0]{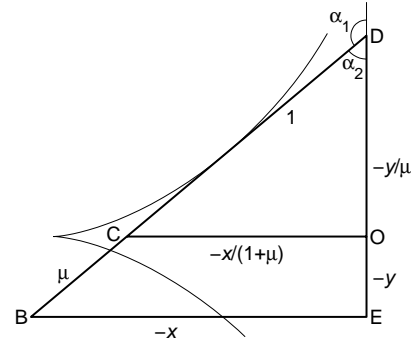}
\end{center}
\caption{\label{astroidfig}
The solution of the astroid equations by similar triangles.  The scaled
coordinates of $B$ are $(x, y)$; $O$ is the point antipodal to $A$. The
line $BCD$, which is given by Eq.~(\ref{anti-line}), is the continuation
of the geodesic from $AB$ with $C$ being its intersection with the
circle $\beta = -\beta_1$ and $D$ its intersection with the meridian
$\lambda = \lambda_1 + \pi$.  The envelope of lines satisfying $CD = 1$
gives the astroid, a portion of which is shown by the curves.}
\end{figure}%
Given $x$ and $y$ (i.e., the position of point $B$),
Eq.~(\ref{anti-line}) may be solved to obtain a first approximation to
$\alpha_1$.  This prescription is given
by \citet[Eq.~(7.3.7)]{helmert80} who notes that this results in a
quartic which may be found using the construction given in
Fig.~\ref{astroidfig}.  Here $COD$ and $BED$ are similar triangles; if
the (signed) length $BC$ is $\mu$, then an equation for $\mu$ can be
found by applying Pythagoras' theorem to $COD$,
\[
\frac{x^2}{(1+\mu)^2} + \frac{y^2}{\mu^2} = 1,
\]
which can be expanded to give a 4th-order polynomial in $\mu$,
\begin{equation}\label{kapeq}
\mu^4 + 2 \mu^3 + (1 - x^2 - y^2) \mu^2 - 2 y^2 \mu - y^2 = 0.
\end{equation}
\begin{table}[tb]
\caption{\label{astroidex}
Second sample inverse calculation specified by $\phi_1 = -30^\circ$, $\phi_2
= 29.9^\circ$, and $\lambda_{12} = 179.8^\circ$.  Because the points are
nearly antipodal, an initial guess for $\alpha_1$ is found by solving
the astroid problem.  Here $\mu$ is the positive root of
Eq.~(\ref{kapeq}).  If $y = 0$, then $\alpha_1$ is given by
Eq.~(\ref{thetaeq0}).  The value of $\alpha_1$ is used in
Table \ref{newtonex}.}
% inversestart(-30,29.9b0,179.8b0);
\begin{center}
\begin{tabular}{@{\extracolsep{0.5\xx}}>{$}c<{$} >{$}l<{$} >{}c<{}}
\hline\hline\noalign{\smallskip}
\text{Qty.} & \hphantom{-}\text{Value} & Eq.
\\\noalign{\smallskip}\hline\noalign{\smallskip}
\phi_1& -30 ^\circ& given \\
\phi_2&\hphantom{0}\signsp 29.9 ^\circ& given \\
\mbox{}\quad\lambda_{12}\quad\mbox{}&\signsp 179.8 ^\circ& given \\
\hline\multicolumn{3}{l}{Solve the astroid problem}\\
x& -0.382\,344 & (\ref{xyeq}) \\
y& -0.220\,189 & (\ref{xyeq}) \\
\mu&\hphantom{-} 0.231\,633 & (\ref{kapeq}) \\
\hline\multicolumn{3}{l}{Initial guess for $\alpha_1$}\\
\alpha_1&\signsp 161.914 ^\circ& (\ref{thetaeq}) \\
\noalign{\smallskip}\hline\hline
\end{tabular}
\end{center}
\end{table}%
Descartes' rule of signs shows that, for $y \ne 0$, there is one
positive root \citep[\S\dlmf{1.11(ii)}{1.11.ii}]{dlmf10} and this is the
solution corresponding to the shortest path.  This root can be found by
standard methods \citep[\S\dlmf{1.11(iii)}{1.11.iii}]{dlmf10}.  Equation
(\ref{kapeq}) arises in converting from geocentric to geodetic
coordinates, and I use the solution to that problem given
by \citet{vermeille02}.  The azimuth can then be determined from the
triangle $COD$ in Fig.~\ref{astroidfig},
\begin{equation}\label{thetaeq}
\alpha_1 = \ph\bigl(y/\mu - i x/(1+\mu)\bigr).
\end{equation}
If $y = 0$, the solution is found by taking the limit $y \rightarrow 0$,
\begin{equation}\label{thetaeq0}
\alpha_1 = \ph\bigl(\pm \sqrt{\max(0, 1 - x^2)} - i x\bigr).
\end{equation}
Tables \ref{astroidex}--\ref{hybridex} together illustrate the complete
solution of the inverse problem for nearly antipodal points.
\begin{table}[tb]
\caption{\label{newtonex}
Second sample inverse calculation, continued.  Here $\lambda_{12}^{(0)}$
denotes the desired value of the longitude difference; Newton's method
is used to adjust $\alpha_1$ so that $\lambda_{12}
= \lambda_{12}^{(0)}$.  The final value of $\alpha_1$ is used in
Table \ref{hybridex}.}
% inverse(-30,29.9b0,179.8b0,161.914b0,0,0);
\begin{center}
\begin{tabular}{@{\extracolsep{0.5\xx}}>{$}c<{$} >{$}l<{$} >{}c<{}}
\hline\hline\noalign{\smallskip}
\text{Qty.} & \hphantom{-}\text{Value} & Eq.
\\\noalign{\smallskip}\hline\noalign{\smallskip}
\phi_1& -30 ^\circ&given\\
\phi_2&\hphantom{-} 29.9 ^\circ&given\\
\alpha_1&\signsp 161.914 ^\circ&Table \ref{astroidex}\\
\lambda_{12}^{(0)}&\signsp 179.8 ^\circ&given\\
\hline\multicolumn{3}{l}{Solve triangle $N\;EA$}\\
\beta_1&\hphantom{0}{ -29.916\,747\,713\,24^\circ }&(\ref{redlat})\\
\alpha_0&\hphantom{-0} 15.609\,397\,464\,14^\circ &(\ref{alp0eq})\\
\sigma_1 & -148.812\,535\,665\,96^\circ &(\ref{sigeq})\\
\omega_1 & -170.748\,966\,961\,28^\circ &(\ref{omgeq})\\
\hline\multicolumn{3}{l}{Solve triangle $N\;EB$}\\
\beta_2&\hphantom{-0} 29.816\,916\,421\,89^\circ &(\ref{redlat})\\
\alpha_2 &\hphantom{-0} 18.067\,287\,962\,31^\circ
                                            &(\ref{napa0a}), (\ref{alp2-eq})\\
\sigma_2 &\hphantom{-0} 31.082\,449\,768\,95^\circ &(\ref{sigeq})\\
\omega_2 &\hphantom{-00} 9.213\,457\,611\,10^\circ &(\ref{omgeq})\\
\hline\multicolumn{3}{l}{Determine $\lambda_{12}$}\\
k^2 &\hphantom{0} 0.006\,251\,537\,916\,62 &(\ref{keq})\\
\epsilon &\hphantom{0} 0.001\,558\,018\,267\,80 &(\ref{epseqa})\\
\lambda_1 & -170.614\,835\,524\,58^\circ &(\ref{lameq})\\
\lambda_2 &\hphantom{-00} 9.185\,420\,098\,39^\circ &(\ref{lameq})\\
\lambda_{12} &\hphantom{-} 179.800\,255\,622\,97^\circ
                                                    &$\lambda_2 - \lambda_1$\\
\hline\multicolumn{3}{l}{Update $\alpha_1$}\\
\delta\lambda_{12} &\hphantom{-} 0.000\,255\,622\,97^\circ
                                        &$\lambda_{12} - \lambda_{12}^{(0)}$\\
J(\sigma_1) & -0.009\,480\,409\,276\,40 &(\ref{jeq})\\
J(\sigma_2) &\hphantom{-} 0.000\,313\,491\,286\,30 &(\ref{jeq})\\
m_{12} &\hphantom{-} 57\,288.000\,110\,\mathrm m &(\ref{meq})\\
\d\lambda_{12}/\d\alpha_1 &\hphantom{0} 0.010\,889\,317\,161\,15
                                                       &(\ref{lambda-deriv})\\
\delta\alpha_1 &\hphantom{0}{ -0.023\,474\,655\,19^\circ
                         }&$-\delta\lambda_{12}/(\d\lambda_{12}/\d\alpha_1)$\\
\alpha_1 &\signsp 161.890\,525\,344\,81^\circ &$\alpha_1 + \delta\alpha_1$\\
\hline\multicolumn{3}{l}{Next iteration}\\
\delta\lambda_{12} &\hphantom{00} 0.000\,000\,006\,63^\circ \\
\alpha_1 & 161.890\,524\,736\,33^\circ &\\
\noalign{\smallskip}\hline\hline
\end{tabular}
\end{center}
\end{table}%
\begin{table}[tb]
\caption{\label{hybridex}
Second sample inverse calculation, concluded.  Here the hybrid problem
($\phi_1$, $\phi_2$, and $\alpha_1$ given) is solved.  The computed value
of $\lambda_{12}$ matches that given in the specification of the inverse
problem in Table \ref{astroidex}.}
% inverse(-30,29.9b0,179.8b0,161.89052473633b0,0,1);
\begin{center}
\begin{tabular}{@{\extracolsep{0.5\xx}}>{$}c<{$} >{$}l<{$} >{}c<{}}
\hline\hline\noalign{\smallskip}
\text{Qty.} & \hphantom{-}\text{Value} & Eq.
\\\noalign{\smallskip}\hline\noalign{\smallskip}
\phi_1& -30 ^\circ&given\\
\phi_2&\hphantom{-} 29.9 ^\circ&given\\
\alpha_1&\signsp 161.890\,524\,736\,33^\circ &Table \ref{newtonex}\\
\hline\multicolumn{3}{l}{Solve triangle $N\;EA$}\\
\beta_1&\hphantom{0}{ -29.916\,747\,713\,24^\circ }&(\ref{redlat})\\
\alpha_0&\hphantom{-0} 15.629\,479\,665\,37^\circ &(\ref{alp0eq})\\
\sigma_1 & -148.809\,136\,917\,76^\circ &(\ref{sigeq})\\
\omega_1 & -170.736\,343\,780\,66^\circ &(\ref{omgeq})\\
\hline\multicolumn{3}{l}{Solve triangle $N\;EB$}\\
\beta_2&\hphantom{-0} 29.816\,916\,421\,89^\circ &(\ref{redlat})\\
\alpha_2 &\hphantom{-0} 18.090\,737\,245\,74^\circ
                                            &(\ref{napa0a}), (\ref{alp2-eq})\\
\sigma_2 &\hphantom{-0} 31.085\,834\,470\,40^\circ &(\ref{sigeq})\\
\omega_2 &\hphantom{-00} 9.226\,028\,621\,10^\circ &(\ref{omgeq})\\
\hline\multicolumn{3}{l}{Determine $s_{12}$ and $\lambda_{12}$}\\
s_1 & -16\,539\,979.064\,227\,\mathrm m &(\ref{disteq})\\
s_2 &\hphantom{-0} 3\,449\,853.763\,383\,\mathrm m &(\ref{disteq})\\
s_{12} &\hphantom{-} 19\,989\,832.827\,610\,\mathrm m &$s_2 - s_1$\\
\lambda_1 & -170.602\,047\,121\,48^\circ &(\ref{lameq})\\
\lambda_2 &\hphantom{-00} 9.197\,952\,878\,52^\circ &(\ref{lameq})\\
\mbox{}\quad\lambda_{12}\quad\mbox{}&\hphantom{-} 179.800\,000\,000\,00^\circ
                                                    &$\lambda_2 - \lambda_1$\\
\hline\multicolumn{3}{l}{Solution}\\
\alpha_1 & 161.890\,524\,736\,33^\circ &\\
\alpha_2 &\hphantom{0} 18.090\,737\,245\,74^\circ &\\
s_{12} & 19\,989\,832.827\,610\,\mathrm m &\\
\noalign{\smallskip}\hline\hline
\end{tabular}
\end{center}
\end{table}

\section{Area}\label{area}

The last geodesic algorithm I present is for geodesic areas.  Here, I
extend the method of \citet{danielsen89} to higher order so that the
result is accurate to round-off, and I recast his series into a simple
trigonometric sum.

Let $S_{12}$ be the area of the geodesic quadrilateral $AF\;H\;B$ in
Fig.~\ref{figtrig}.  Following \citet{danielsen89}, this can be
expressed as the sum of a spherical term and an integral giving the
ellipsoidal correction,
\begin{align}
S_{12} &= S(\sigma_2) - S(\sigma_1),\label{S12}\displaybreak[0]\\
S(\sigma) &= c^2\alpha + e^2a^2\cos\alpha_0 \sin\alpha_0 \,I_4(\sigma),
\label{Sdef}
\end{align}
where
\begin{equation}\label{authalic}
c^2 = \frac{a^2}2 + \frac{b^2}2 \frac{\tanh^{-1}e}e
\end{equation}
is the authalic radius,
\begin{align}
I_4(\sigma) &= -\int_{\pi/2}^\sigma
\frac{t(e'^2) - t(k^2\sin^2\sigma')}{e'^2-k^2\sin^2\sigma'}
\frac{\sin\sigma'}2 \,\d\sigma',
\label{I4eq}\\
t(x) &= x + \sqrt{x^{-1} + 1}\,\sinh^{-1}\!\sqrt x.\notag
\end{align}
Expanding the integrand in powers of $e'^2$ and $k^2$ and performing the
integral gives
\begin{equation}\label{i4}
I_4(\sigma) = \sum_{l = 0}^\infty C_{4l}\cos \bigl((2l+1)\sigma\bigr),
\end{equation}
where
\begin{align} \label{C4}
C_{40} &= \bigl(\tfrac2/3 - \tfrac1/15 e'^2 + \tfrac4/105 e'^4
- \tfrac8/315 e'^6 + \tfrac64/3465 e'^8 - \tfrac128/9009
e'^{10}\bigr)\notag\\&\quad - \bigl(\tfrac1/20 - \tfrac1/35 e'^2
+ \tfrac2/105 e'^4 - \tfrac16/1155 e'^6 + \tfrac32/3003 e'^8 \bigr)
k^2\notag\\&\quad + \bigl(\tfrac1/42 - \tfrac1/63 e'^2 + \tfrac8/693
e'^4 - \tfrac80/9009 e'^6\bigr) k^4\notag\\&\quad - \bigl(\tfrac1/72
- \tfrac1/99 e'^2 + \tfrac10/1287 e'^4 \bigr) k^6\notag\\&\quad
+ \bigl(\tfrac1/110 - \tfrac1/143 e'^2\bigr) k^8 - \tfrac1/156 k^{10}
+ \cdots,\displaybreak[0]\notag\\
C_{41} &= \bigl(\tfrac1/180 - \tfrac1/315 e'^2 + \tfrac2/945 e'^4
- \tfrac16/10395 e'^6 + \tfrac32/27027 e'^8\bigr) k^2\notag\\&\quad
- \bigl(\tfrac1/252 - \tfrac1/378 e'^2 + \tfrac4/2079 e'^4
- \tfrac40/27027 e'^6\bigr) k^4\notag\\&\quad + \bigl(\tfrac1/360
- \tfrac1/495 e'^2 + \tfrac2/1287 e'^4 \bigr) k^6\notag\\&\quad
- \bigl(\tfrac1/495 - \tfrac2/1287 e'^2\bigr) k^8 + \tfrac5/3276 k^{10}
+ \cdots,\displaybreak[0]\notag\\
C_{42} &= \bigl(\tfrac1/2100 - \tfrac1/3150 e'^2 + \tfrac4/17325
e'^4 - \tfrac8/45045 e'^6\bigr) k^4\notag\\&\quad - \bigl(\tfrac1/1800
- \tfrac1/2475 e'^2 + \tfrac2/6435 e'^4 \bigr) k^6\notag\\&\quad
+ \bigl(\tfrac1/1925 - \tfrac2/5005 e'^2\bigr) k^8 - \tfrac1/2184 k^{10}
+ \cdots,\displaybreak[0]\notag\\
C_{43} &= \bigl(\tfrac1/17640 - \tfrac1/24255 e'^2 + \tfrac2/63063
e'^4 \bigr) k^6\notag\\&\quad - \bigl(\tfrac1/10780 - \tfrac1/14014
e'^2\bigr) k^8 + \tfrac5/45864 k^{10} + \cdots,\displaybreak[0]\notag\\
C_{44} &= \bigl(\tfrac1/124740 - \tfrac1/162162 e'^2\bigr) k^8
- \tfrac1/58968 k^{10} + \cdots,\displaybreak[0]\notag\\
C_{45} &= \tfrac1/792792 k^{10} + \cdots.
\end{align}
\begin{table}[tb]
\caption{\label{areaex}
The calculation of the area between the equator and the geodesic
specified by $\phi_1 = 40^\circ$, $\alpha_1 = 30^\circ$, and $s_{12} =
10\,000\,\mathrm{km}$.  This uses intermediate values computed in
Table \ref{directex}.}
% directa(40, 30, 10b6);
\begin{center}
\begin{tabular}{@{\extracolsep{0.5\xx}}>{$}c<{$} >{$}l<{$} >{}c<{}}
\hline\hline\noalign{\smallskip}
\text{Qty.} & \text{Value} & Eq.
\\\noalign{\smallskip}\hline\noalign{\smallskip}
\alpha_0 &\hphantom{0} 22.553\,940\,202\,62^\circ &Table \ref{directex}\\
\alpha_1 &\hphantom{0} 30^\circ &Table \ref{directex}\\
\alpha_2 & 149.090\,169\,318\,07^\circ &Table \ref{directex}\\
\sigma_1 &\hphantom{0} 43.999\,153\,645\,00^\circ &Table \ref{directex}\\
\sigma_2 & 133.921\,640\,830\,38^\circ &Table \ref{directex}\\
k^2 & 0.005\,748\,029\,628\,57 &Table \ref{directex}\\
\hline\multicolumn{3}{l}{Compute area}\\
I_4(\sigma_1) &\hphantom{-} 0.479\,018\,145\,20 &(\ref{i4})\\
I_4(\sigma_2) & -0.461\,917\,119\,02 &(\ref{i4})\\
S(\sigma_1) &\hphantom{0} 21\,298\,942.667\,15\,\mathrm{km}^2 &(\ref{Sdef})\\
S(\sigma_2) & 105\,574\,566.089\,50\,\mathrm{km}^2 &(\ref{Sdef})\\
S_{12} &\hphantom{0} 84\,275\,623.422\,35\,\mathrm{km}^2 &(\ref{S12})\\
\noalign{\smallskip}\hline\hline
\end{tabular}
\end{center}
\end{table}%
An example of the computation of $S_{12}$ is given in Table \ref{areaex}.

Summing $S_{12}$, Eq.~(\ref{S12}), over the edges of a geodesic polygon
gives the area of the polygon provided that it does not encircle a pole;
if it does, $2\pi c^2$ should be added to the result.  The first term in
Eq.~(\ref{Sdef}) contributes $c^2(\alpha_2 - \alpha_1)$ to $S_{12}$.
This is the area of the quadrilateral $AF\;H\;B$ on a sphere of radius
$c$ and it is proportional to its spherical excess, $\alpha_2
- \alpha_1$, the sum of its interior angles less $2\pi$.  It is
important that this term be computed accurately when the edge is short
(and $\alpha_1$ and $\alpha_2$ are nearly equal).  A suitable identity
for $\alpha_2 - \alpha_1$ is given by \citet[\S11]{bessel25},
\begin{equation}
\tan\frac{\alpha_2 - \alpha_1}2 =
\frac{\sin\tfrac1/2 (\beta_2 + \beta_1)}
{\cos\tfrac1/2 (\beta_2 - \beta_1)} \tan\frac{\omega_{12}}2.
\end{equation}

\section{Implementation}\label{implement}

The algorithms described in the preceding sections can be readily
converted into working code.  The polynomial expansions,
Eqs.~(\ref{A1}), (\ref{C1}), (\ref{C11}), (\ref{A3}), (\ref{C3}),
(\ref{A2}), (\ref{C2}), and (\ref{C4}), are such that the final results
are accurate to $O(f^6)$ which means that, even for $f = \frac1{150}$,
the truncation error is smaller than the round-off error when using IEEE
double precision arithmetic (with the fraction of the floating point
number represented by 53 bits).  For speed and to minimize round-off
errors, the polynomials should be evaluated with the Horner method.  The
parenthetical expressions in Eqs.~(\ref{A3}), (\ref{C3}), and (\ref{C4})
depend only on the flattening of the ellipsoid and can be computed once
this is known.  When determining many points along a single geodesic,
the polynomials need be evaluated just once.  \citet{clenshaw55}
summation should be used to sum the Fourier series, Eqs.~(\ref{i1expr}),
(\ref{i3expr}), (\ref{i2expr}), and (\ref{i4}).

There are several other details to be dealt with in implementing the
algorithms: where to apply the two rules for choosing starting points
for Newton's method, a slight improvement to the starting guess
Eq.~(\ref{thetaeq}), the convergence criterion for Newton's method, how
to minimize round-off errors in solving the trigonometry problems on the
auxiliary sphere, rapidly computing intermediate points on a geodesic by
using $\sigma_{12}$ as the metric, etc.  I refer the reader to the
implementations of the algorithms in
GeographicLib \citep{geographiclib120} for possible ways to address
these issues.  The C++ implementation has been tested against a
large set of geodesics for the WGS84 ellipsoid; this was generated by
continuing the series expansions to $O(f^{30})$ and by solving the
direct problem using with high-precision arithmetic.  The round-off
errors in the direct and inverse methods are less than 15 nanometers
and the error in the computation of the area $S_{12}$ is about
$0.1\,\mathrm{m^2}$.  Typically, 2 to 4 iterations of Newton's method
are required for convergence, although in a tiny fraction of cases up to
16 iterations are required.  No convergence failures are observed.  With
the C++ implementation compiled with the g++ compiler, version 4.4.4,
and running on a $2.66\,\mathrm{GHz}$ Intel processor, solving the
direct geodesic problem takes $0.88\,\mathrm{\mu s}$, while the inverse
problem takes $2.34\,\mathrm{\mu s}$ (on average).  Several points along
a geodesic can be computed at the rate of $0.37\,\mathrm{\mu s}$ per
point.  These times are comparable to those for Vincenty's algorithms
implemented in C++ and run on the same architecture: $1.11\,\mathrm{\mu
s}$ for the direct problem and $1.34\,\mathrm{\mu s}$ for the inverse
problem.  (But note that Vincenty's algorithms are less accurate than
those given here and that his method for the inverse problem sometimes
fails to converge.)

\section{Ellipsoidal gnomonic projection}\label{gnomproj}

As an application of the differential properties of geodesics, I derive
a generalization of the gnomonic projection to the ellipsoid.  The
gnomonic projection of the sphere has the property that all geodesics on
the sphere map to straight lines \citep[\S22]{snyder87}.  Such a
projection is impossible for an ellipsoid because it does not have
constant Gaussian curvature \citep[\S18]{beltrami65}; nevertheless, a
projection can be constructed in which geodesics are very nearly
straight.

\begin{figure}[tb]
\begin{center}
\includegraphics[scale=0.75,angle=0]{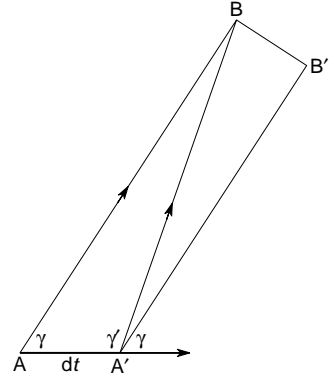}
\end{center}
\caption{\label{gnomconstr}
The construction of the generalized gnomonic projection as the limit of
a doubly azimuthal projection.}
\end{figure}

The spherical gnomonic projection is the limit of the doubly azimuthal
projection of the sphere, wherein the bearings from two fixed points $A$
and $A'$ to $B$ are preserved, as $A'$ approaches
$A$ \citep{bugayevskiy95}.  The construction of the generalized gnomonic
projection proceeds in the same way; see Fig.~\ref{gnomconstr}.  Draw a
geodesic $A'B'$ such that it is parallel to the geodesic $AB$ at $A$.
Its initial separation from $AB$ is $\sin\gamma\,\d t$; at $B'$, the
point closest to $B$, the separation becomes $M_{12}\sin\gamma\,\d t$
(in the limit $\d t\rightarrow 0$).  Thus the difference in the azimuths
of the geodesics $A'B$ and $A'B'$ at $A'$ is
$(M_{12}/m_{12})\sin\gamma\,\d t$, which gives $\gamma + \gamma' = \pi -
(M_{12}/m_{12})\sin\gamma\,\d t$.  Now, solving the planar triangle
problem with $\gamma$ and $\gamma'$ as the two base angles gives the
distance $AB$ on the projection plane as $m_{12}/M_{12}$.

This leads to the following specification for the generalized gnomonic
projection.  Let the center point be $A$; for an arbitrary point $B$,
solve the inverse geodesic problem between $A$ and $B$; then $B$
projects to the point
\begin{equation}\label{gnom-eq}
x = \rho\sin\alpha_1, \quad y = \rho\cos\alpha_1, \quad
\rho = m_{12}/M_{12};
\end{equation}
the projection is undefined if $M_{12} \le 0$.  In the spherical limit,
this becomes the standard gnomonic projection, $\rho =
a \tan\sigma_{12}$ \citep[p.~165]{snyder87}.  The azimuthal scale is
$1/M_{12}$ and the radial scale, found by taking the derivative
$\d\rho/\d s_{12}$ and using Eq.~(\ref{wronski}), is $1/M_{12}^2$.  The
reverse projection is found by computing $\alpha_1 = \ph(y + ix)$,
finding $s_{12}$ using Newton's method with $\d\rho/\d s_{12} =
1/M_{12}^2$ (i.e., the radial scale), and solving the resulting direct
geodesic problem.

In order to gauge the usefulness of the ellipsoidal gnomonic projection,
consider two points on the earth $B$ and $C$, map these points to the
projection, and connect them with a straight line in this projection.
If this line is mapped back onto the surface of the earth, it will
deviate slightly from the geodesic $BC$.  To lowest order, the maximum
deviation $\v h$ occurs at the midpoint of the line segment $BC$;
empirically, I find
\begin{equation}\label{gnom-h}
\v h = \frac{l^2}{32} (\nabla K \cdot \v t) \v t,
\end{equation}
where $l$ is the length of the geodesic, $K$ is the Gaussian curvature,
$\nabla K$ is evaluated at the center of the projection $A$,
and $\v t$ is the perpendicular vector from the center of projection to
the geodesic.  The deviation in the azimuths at the end points
is about $4h/l$ and the length is greater than the geodesic distance by
about $\frac83 h^2/l$.  In the case of an ellipsoid of revolution, the
curvature is given by differentiating Eq.~(\ref{curvature}) with respect
to $\phi$ and dividing by the meridional radius of curvature to give
\begin{equation}
\nabla K = -\frac{4a}{b^4}
e^2(1-e^2\sin^2\phi)^{5/2}\cos\phi\sin\phi\, \v{\hat {\boldsymbol\phi}},
\end{equation}
where $\v{\hat {\boldsymbol\phi}}$ is a unit vector pointing north.
Bounding the magnitude of $\v h$, Eq.~(\ref{gnom-h}), over all the
geodesics whose end points lie within a distance $r$ of the center of
projection, gives (in the limit that $f$ and $r$ are small)
\begin{equation}\label{gnomonic-err}
\frac hr \le \frac f8 \frac{r^3}{a^3}.
\end{equation}
The maximum value is attained when the center of projection is at $\phi
 = \pm 45^\circ$ and the geodesic is running in an east-west direction
 with the end points at bearings $\pm 45^\circ$ or $\pm 135^\circ$ from
 the center.

\begin{figure}[tb]
\begin{center}
\includegraphics[scale=0.75,angle=0]{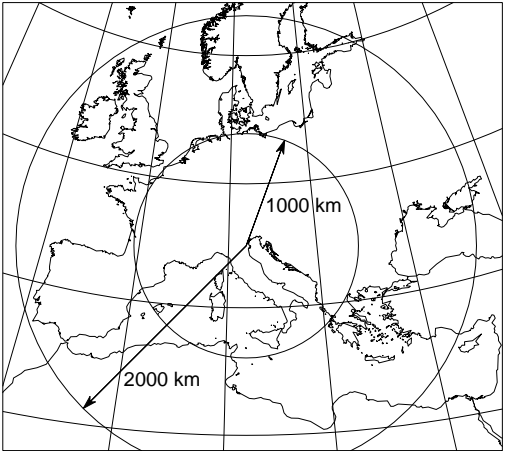}
\end{center}
\caption{\label{gnomfig}
The coast line of Europe and North Africa in the ellipsoidal gnomonic
projection with center at $(45^\circ\mathrm N, 12^\circ\mathrm E)$ near
Venice.  The graticule lines are shown at multiples of $10^\circ$.  The
two circles are centered on the projection center with (geodesic) radii
of $1000\,\mathrm{km}$ and $2000\,\mathrm{km}$.  The data for the coast
lines is taken from GMT \citep{gmt455} at ``low'' resolution.}
\end{figure}%
Others have proposed different generalizations of the gnomonic
projection.  \citet{bowring97} and \citet{williams97} give a projection
in which great ellipses project to straight lines; \citet{letovaltsev63}
suggests a projection in which normal sections through the center point
map to straight lines.  Empirically, I find that $h/r$ is proportional
to $r/a$ and $r^2/a^2$ for these projections.  Thus, neither does as
well as the projection derived above (for which $h/r$ is proportional to
$r^3/a^3$) at preserving the straightness of geodesics.

As an illustration of the properties of the ellipsoidal gnomonic
projection, Eq.~(\ref{gnom-eq}), consider Fig.~\ref{gnomfig} in which a
projection of Europe is shown.  The two circles are geodesic circles of
radii $1000\,\mathrm{km}$ and $2000\,\mathrm{km}$.  If the geodesic
between any two points within one of these circles is estimated by using
a straight line on this figure, the deviation from the true geodesic is
less than $1.7\,\mathrm m$ and $28\,\mathrm m$, respectively.  The
maximum errors in the end azimuths are $1.1''$ and $8.6''$ and the
maximum errors in the lengths are only $5.4\,\mu\mathrm m$ and
$730\,\mu\mathrm m$.

The gnomonic projection can be used to solve two geodesic problems
accurately and rapidly.  The first is the {\it intersection} problem:
given two geodesics between $A$ and $B$ and between $C$ and $D$,
determine the point of intersection, $O$.  This can be solved as
follows.  Guess an intersection point $O^{(0)}$ and use this as the
center of the gnomonic projection; define $\v a$, $\v b$, $\v c$, $\v d$
as the positions of $A$, $B$, $C$, $D$ in the projection; find the
intersection of $AB$ and $CD$ in the projection, i.e.,
\begin{equation}
\v o = \frac
{ (\v c \times \v d \cdot \v{\hat z}) (\v b - \v a)
- (\v a \times \v b \cdot \v{\hat z}) (\v d - \v c)}
{(\v b - \v a) \times (\v d - \v c) \cdot \v{\hat z}},
\end{equation}
where $\v{\hat{\text{\ }}}$ indicates a unit vector ($\v{\hat a} = \v
a/a$) and $\v{\hat z} = \v{\hat x} \times \v{\hat y}$ is in the
direction perpendicular to the projection plane.  Project $\v o$ back to
geographic coordinates $O^{(1)}$ and use this as a new center of
projection; iterate this process until $O^{(i)} = O^{(i-1)}$ which is
then the desired intersection point.

The second problem is the {\it interception} problem: given a geodesic
between $A$ and $B$, find the point $O$ on the geodesic which is closest
to a given point $C$.  The solution is similar to that for the
intersection problem; however the interception point in the projection
is
\[
\v o = \frac
{\v c \cdot (\v b - \v a) (\v b - \v a)
- (\v a \times \v b \cdot \v{\hat z}) \v{\hat z} \times (\v b - \v a)}
{\abs{\v b - \v a}^2}.
\]
Provided the given points lie within about a quarter meridian of the
intersection or interception points (so that the gnomonic projection is
defined), these algorithms converge quadratically to the exact result.

\section{Conclusions}\label{conclusions}

The classical geodesic problems entail solving the ellipsoidal triangle
$N\;AB$ in Fig.~\ref{figtrig}, whose sides and angles are represented by
$\phi_1$, $\phi_2$, $s_{12}$ and $\alpha_1$, $\alpha_2$, $\lambda_{12}$.
In the direct problem $\phi_1$, $\alpha_1$, and $s_{12}$ are given,
while in the inverse problem $\phi_1$, $\lambda_{12}$, and $\phi_2$ are
specified; and the goal in each case is to solve for the remaining side
and angles.  The algorithms given here provide accurate, robust, and
fast solutions to these problems; they also allow the differential and
integral quantities $m_{12}$, $M_{12}$, $M_{21}$, and $S_{12}$ to be
computed.

Much of the work described here involves applying standard computational
techniques to earlier work.  However, at least two aspects are novel:
(1)~This paper presents the first complete solution to the inverse
geodesic problem.  (2)~The ellipsoidal gnomonic projection is a new tool
to solve various geometrical problems on the ellipsoid.

Furthermore, the packaging of these various geodesic capabilities into a
single library is also new.  This offers a straightforward solution of
several interesting problems.  Two geodesic projections, the azimuthal
equidistant projection and the Cassini-Soldner projection, are simple to
write and their domain of applicability is not artificially restricted,
as would be the case, for example, if the series expansion for the
Cassini-Soldner projection were used \citep[\S13]{snyder87}; the scales
for these projections are simply given in terms of $m_{12}$ and
$M_{12}$.  Several other problems can be readily tackled with this
library, e.g., solving other ellipsoidal trigonometry problems and
finding the median line and other maritime boundaries.  These and other
problems are explored in \citet{karney11a}.  The web
page \url{http://geographiclib.sf.net/geod.html} provides additional
information, including the \citet{maxima} code used to carry out the
Taylor expansions and a JavaScript implementation which allows geodesic
problems to be solved on many portable devices.

\section*{Acknowledgments}

I would like to thank Rod Deakin, John Nolton, Peter Osborne, and the
referees of this paper for their helpful comments.

\bibliography{geod}
\end{document}